\documentclass[10pt,letterpaper]{article}
\usepackage{opex3}
\usepackage{color}
\usepackage{amsmath,amssymb,graphicx,bm,float}
\usepackage{cite}

\newcommand{\ket}[1]{|{#1}\rangle}

\def\ts#1{{\textstyle #1}}

\def\onehalf{{\textstyle\frac12}}

\def\'#1{\if#1i{\accent 19\i}\else{\accent 19 #1}\fi}

\begin{document}
\title{Heralded single-photon source utilizing highly nondegenerate, spectrally factorable spontaneous parametric downconversion}
\author{Fumihiro Kaneda,$^1$ Karina Garay-Palmett,$^2$ Alfred B. U'Ren,$^3$ and Paul G. Kwiat$^1$}
\address{$^1$Department of Physics, University of Illinois, Urbana-Champaign, IL 61801, USA \\
$^2$ Departamento de {\'O}ptica, Centro de Investigaci{\'o}n Cient{\'i}fica y de Educaci{\'o}n Superior de Ensenada, Apartado Postal 360 Ensenada, BC 22860, Mexico\\ 
$^3$Instituto de Ciencias Nucleares, Universidad Nacional Aut{\'o}noma de M{\'e}xico, apdo. postal 70-543, 04510 D.F., Mexico}
\date{\today}

\begin{abstract}
We report on the generation of an indistinguishable heralded single-photon state, using highly nondegenerate spontaneous parametric downconversion (SPDC). 
Spectrally factorable photon pairs can be generated by incorporating a broadband pump pulse and a group-velocity matching (GVM) condition in a periodically-poled potassium titanyl phosphate (PPKTP) crystal. 
The heralding photon is in the near IR, close to the peak detection efficiency of off-the-shelf Si single-photon detectors; meanwhile, the heralded photon is in the telecom L-band where fiber losses are at a minimum. 
We observe spectral factorability of the SPDC source and consequently high purity (90\%) of the produced heralded single photons by several different techniques. 
Because this source can also realize a high heralding efficiency ($>90$\%), it would be suitable for time-multiplexing techniques, enabling a pseudo-deterministic single-photon source, a critical resource for optical quantum information and communication technology. 
\end{abstract}

\bibliographystyle{osajnl.bst}

\section{Introduction}
Single-photon sources are critical resources for quantum-enhanced technologies, such as quantum metrology \cite{Giovannetti:2006ud}, quantum communications \cite{Gisin:2002gb}, and photonic quantum computation \cite{Knill:2001is,OBrien:2007io,Childs:2013hh}. 
Heralded single-photon sources (HSPSs), first demonstrated by Hong and Mandel \cite{Hong:1986va} in 1986, have been of great interest in quantum information applications. 
Heralded single photons may be created by a photon-pair producing nonlinear optical process such as spontaneous parametric downconversion (SPDC) or spontaneous four-wave mixing (SFWM); a single photon's presence is ``heralded'' by the detection of its twin photon. 
Although HSPSs are not on-demand sources due to the probabilistic nature of the pair generation process, they are easier to design and implement compared to single emitter sources \cite{Eisaman:2011cc} such as single atoms, ions, and semiconductor devices, which typically require vacuum and/or cryogenic systems. 
Moreover, it is possible to improve a HSPS to be pseudo-on-demand by using temporal or spatial multiplexing techniques \cite{Pittman:2002dx,migdall:2002hk,Kaneda:2015dn,Mendoza:2016gr}.

In the context of quantum information applications, there are essentially three key metrics for an HSPS: indistinguishability, heralding efficiency, and source brightness. 
Indistinguishable photons, i.e., pure and identical photons, are desirable because they can exhibit multi-photon interference \cite{Hong:1987vi} that is central to many quantum information applications. 
In order to permit heralding of pure single photons, the SPDC or SFWM two-photon state source must be factorable. 
Heralding efficiency, the probability of a single-photon state, conditional on a heralding detector signal, corresponds to the purity of a heralded state in the photon-number basis, and thus is a very important factor to scale up quantum information protocols. 
Lastly, source brightness is often characterized as the photon generation rate per unit pump power (e.g., cps/mW). 
A bright source can relax pump power requirements, and lead, e.g., to higher success probability for multi-photon protocols. 
It is a challenging task to optimize these three metrics simultaneously, as there are usually trade-offs between them.  
For example, a typical SPDC source does not generate spectrally factorable photon pairs and requires narrowband spectral filtering for high purity and indistinguishability. 
However, the spectral filtering significantly reduces source brightness and heralding efficiency.

In order to realize a practical high-quality HSPS, several experiments \cite{Pelton:2004uk,Huebel:2007tg,Hentschel:2009ej,Pomarico:2012uj,Stuart:2013kj,Krapick:2013up} have demonstrated highly nondegenerate SPDC. 
For high brightness of heralded single photons, it is important to use not only a highly efficient SPDC source but also efficient trigger detectors. 
In this sense, it is advantageous to generate trigger photons within the range of the high sensitivity wavelength (600-800 nm) for off-the-shelf, efficient Si avalanche photodiodes (Si-APDs). 
On the other hand, for high total system heralding efficiency, it is preferable to generate photons in the telecommunication band for lower-loss propagation through optical fibers. 
Thus, such nondegenerate SPDC is useful to achieve high brightness and heralding efficiency in practice. 
However, to the best of our knowledge, until now there has been no source based on highly nondegenerate SPDC that simultaneously achieved high purity, due to the presence of strong spectral correlations between the photons. 

Here we propose and demonstrate an HSPS based on highly nondegenerate, spectrally factorable SPDC which can simultaneously attain high purity, high heralding efficiency, and a moderately high brightness. 
A group velocity matching (GVM) condition in a periodically-poled potassium titanyl phosphate (PPKTP) crystal allows us to produce intrinsically factorable photon pairs at $\sim$800 nm and $\sim$1590 nm, which is in the telecommunication L-band. 
Our spectrally factorable SPDC source can generate pure heralded single photons without the narrowband filtering which typically reduces the brightness and heralding efficiency. 
We characterize the purity of the heralded single photons using several techniques, which consistently reveal high state purity. 
Finally, we discuss the possibility of further improvements to our source.

\section{Highly nondegenerate SPDC with GVM condition}

\begin{tiny}
\begin{figure}[t!]
  \includegraphics[width=1.0\columnwidth,clip]{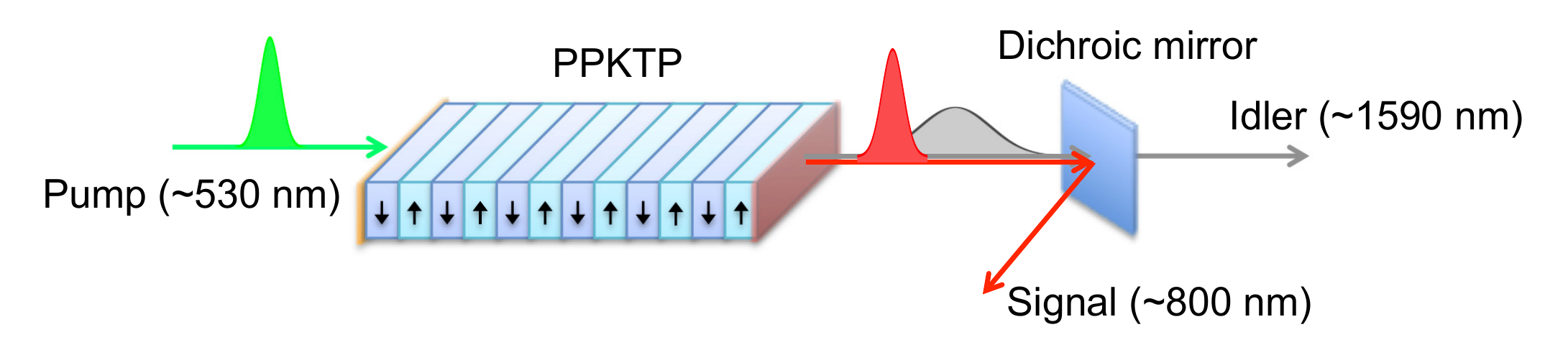}
   \caption{Schematic representation of an SPDC source using a PPKTP crystal.  }
\label{PPKTP}
\end{figure}
\end{tiny}

The basic configuration of our HSPS is depicted in Fig. \ref{PPKTP}. 
Laser pulses with a wavelength of $\sim$530 nm are used to pump a PPKTP crystal with crystal length $L$ and poling period $\Lambda$, designed so that it collinearly produces signal-idler photon pairs at $\sim$800 nm and $\sim$1590 nm in order to obtain practical benefits of higher detection efficiency by Si-APDs (at $\sim$ 800 nm) and high optical-fiber transmission (at $\sim$1590 nm). 
In general, SPDC has the highest conversion efficiency when the pump, signal, and idler modes satisfy energy and momentum conservation (i.e., phase-matching condition): 
\begin{equation}
 \Delta \omega  =\omega_p - \omega_s -  \omega_i = 0, 
\label{energyconservation}
\end{equation}
\begin{equation}
\Delta k    = k_p- k_s -  k_i + \frac{2\pi} {\Lambda}= 0, 
\label{phasematching}
\end{equation}
where $\omega_j$ and $k_j$ are angular frequency and the wavenumber of the pump ($j = p$), signal ($j = s$), and idler ($j = i$) modes, respectively. 
For simplicity, we assume here that the fields which participate in the SPDC process are in the form of plane waves; in Appendix B we present a version of the two-photon state, where the pump is a more realistic Gaussian beam, and where the signal and idler photons are collected by Gaussian-beam modes, leading to single-mode optical fibers (SMFs). 
Under the plane-wave approximation for the three interacting waves, the two-photon joint spectral state (omitting a normalizing factor) is given by 
\begin{align}
\label{statevector}
 \ket{\psi_{si}} =  \int d\omega_s d\omega_i  f(\omega_s, \omega_i ) \ket{\omega_s, \omega_i}, 
\end{align}
where 
$f (\omega_s, \omega_i )$ represents the joint spectral amplitude (JSA), 
and  $\ket{\omega_s, \omega_i}$ denotes a photon-pair state with signal and idler frequencies $\omega_s$ and $\omega_i$, respectively. 
For the specifics of our experiment involving collinear downconversion with a pump which is not too tightly focused, the JSA is well approximated by $f(\omega_s, \omega_i ) \simeq  \alpha (\omega_s, \omega_i ) \phi(\Delta k L)$. 
Here, $\alpha(\omega_s, \omega_i )$ is the pump spectral envelope function, assumed to be well described by a Gaussian function: 
\begin{align}
\alpha (\omega_s, \omega_i ) &= \exp\left[{-\frac{(\omega_s + \omega_i -\omega_{p0} )^2}{\sigma_p^2}} \right] \\
                                                 &= \exp\left[{-\frac{(\Omega_s + \Omega_i )^2}{\sigma_p^2}} \right],
\label{pumpenvelope}
\end{align}
where $\Omega_{s(i)}$ is the frequency detuning from the signal (idler) central frequency $\omega_{s0(i0)}$: $\Omega_{s(i)} =  \omega_{s(i)} -\omega_{s0(i0)}$, and $\omega_{p0} =  \omega_{s0} +\omega_{i0}$. 
$\sigma_p$ denotes the pump spectral bandwidth. 
The joint distribution $\alpha (\Omega_s, \Omega_i )$ has a maximum value when energy-conservation condition shown in Eq. \eqref{energyconservation} is satisfied: $\Omega_i/ \Omega_s = -1$. 
The phase-matching function $\phi(\Delta k L)$ in a nonlinear crystal is given by
\begin{align}
 \phi(\Delta k L) = \mathrm{sinc} \left( \frac{\Delta k L}{2}\right) = \frac{\sin \left( \Delta k L/2\right)}{\Delta k L/2}. 
  \label{phasematchingenvelope}
\end{align}
While $\alpha (\Omega_s, \Omega_i )$ is maximized according to the energy conservation, the distribution $\phi(\Delta k L)$ is governed by dispersion in a nonlinear crystal. 
In our highly nondegenerate SPDC, $\Delta k$ can be approximated by the first-order dispersion: 
\begin{align}
 \Delta k(\Omega_s, \Omega_i ) \simeq k'_p (\Omega_s + \Omega_i )  - k'_s\Omega_s  - k'_i \Omega_i,  
 \label{dkapproximation}
\end{align}
where $k'_j \equiv \frac{d k_j}{d \omega_j}|_{\omega_{j0}}$ is  the inverse group velocity.  
The phase-matching function has its maximum value for the case that $\Omega_i/\Omega_s$ is equal to the dispersion parameter $D = - (k'_p-k'_s)/(k'_p-k'_i)$. 
Thus, in order to obtain a desirable (i.e. factorable) JSA, one can choose appropriate parameters $\sigma_p$, $L$, and $D$. 

\begin{tiny}
\begin{figure}[t!]
\begin{center}
  \includegraphics[width=1\columnwidth,clip]{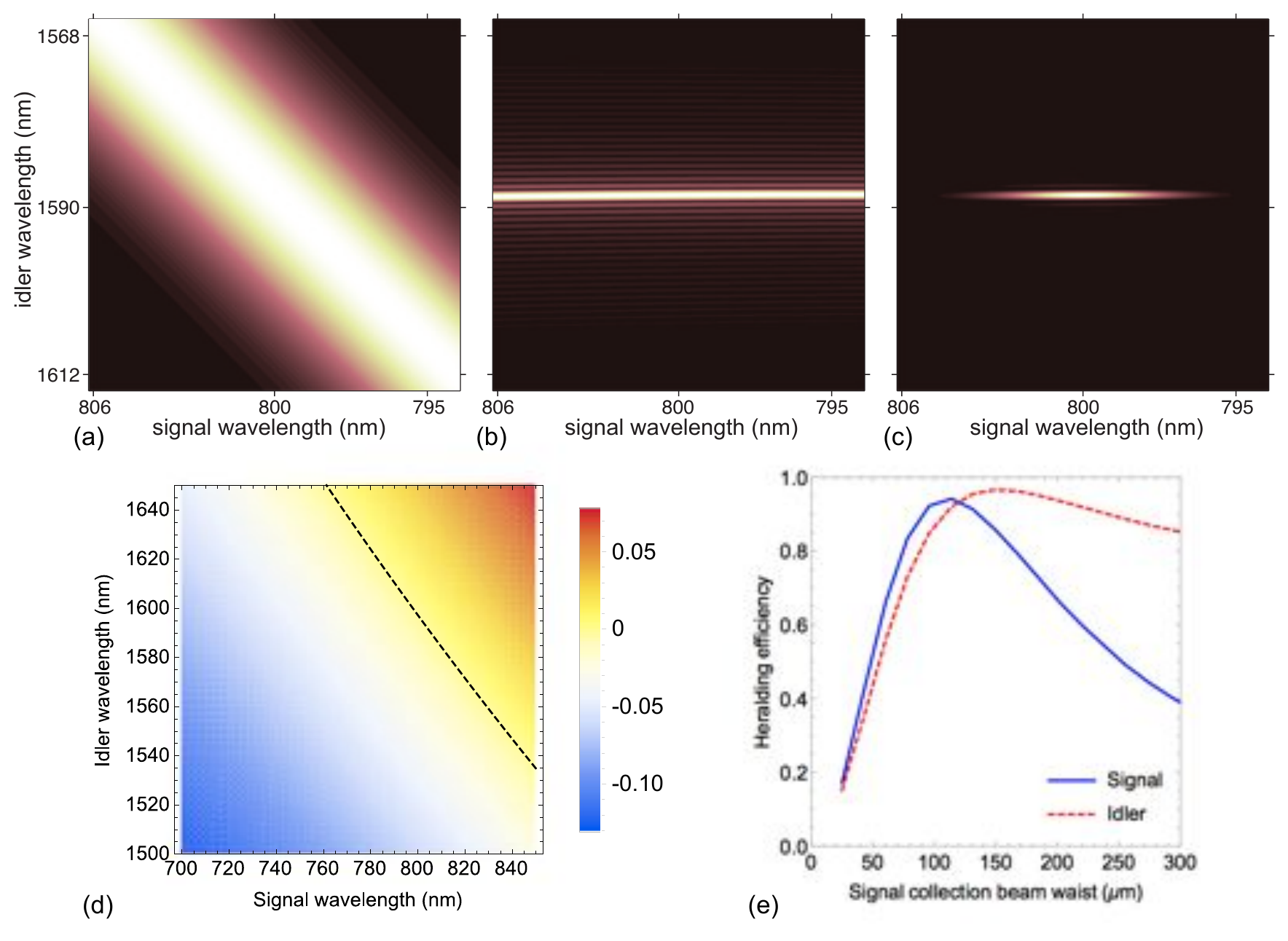}
   \caption{Theoretical predictions. (a) Pump spectral envelope function $\alpha (\Omega_s, \Omega_i )$; (b) Phase-matching function $\phi(\Delta k (\Omega_s, \Omega_i ) L)$, projected to the fiber collection modes (see Appendix B); (c) Fiber-coupled JSI $|f(\Omega_s, \Omega_i )|^2$; (d) Dispersion parameter $D = - (k'_p-k'_s)/(k'_p-k'_i)$; the dashed line shows where $D = 0$. (e) Heralding efficiencies versus signal collection beam waist for pump beam and idler collection beam waists of 220 $\mu$m and 120 $\mu$m, respectively.  }
\label{predictions}
\end{center}
\end{figure}
\end{tiny}

Here we show how to make a factorable JSA, i.e., so that it may be written as $f(\Omega_s, \Omega_i ) = A(\Omega_s)B(\Omega_i)$ in order to generate a pure heralded single-photon state.  
We use Type-II phase-matching: pump, signal, and idler fields are polarized respectively along the $Y$, $Z$, and $Y$ crystal axes for a PPKTP crystal. 
(Note that while our simulations incorporate the full bi-axial angle-dependence of the refractive indices, because the crystal used in our experiment is cut parallel to the crystallographic axes, with propagation of the three waves along the $X$ direction, the signal and idler refractive indices are approximately equal to $n_Z$ and $n_Y$, respectively).  
Figure \ref{predictions} (d) shows $D$ plotted as a function of the  signal and idler wavelengths. 
We can see that $D$ is approximately zero for wide ranges of signal and idler wavelengths around 800 nm and 1590 nm, respectively. 
For $D = 0$, one can obtain a factorable JSA by adjusting the pump bandwidth to be much larger than the phase-matching bandwidth  \cite{URen:2005wb}: $\sigma_p \gg 2/ (k'_p-k'_i) L\sqrt{ \gamma}$, where $\gamma = 0.193$ is a coefficient used for the approximation $\mathrm{sinc} (x) \simeq  \exp (-\gamma x^2)$. 
Figure \ref{predictions} (a-c) show examples of $\alpha (\Omega_s, \Omega_i )$, $\phi(\Delta k (\Omega_s, \Omega_i ) L)$ , and joint spectral intensity (JSI) $|f(\Omega_s, \Omega_i )|^2$ for $\sigma_p =$5 THz, $L = 20$ mm, and $2/ (k'_p-k'_i) L\sqrt{ \gamma}  = $ 0.4 THz. 
Note that our simulations presented in Fig. \ref{predictions} take into account the projection of the two-photon joint amplitude on the fiber-collection modes (see Appendix B). 
We can clearly see the factorability of $|f(\Omega_s, \Omega_i )|^2$ shown in Fig. \ref{predictions} (c). 
Using Schmidt decomposition \cite{Lamata:2005tn}, we estimate the expected attainable purity with this SPDC source to be 97\%.

In terms of attainable purity, utilizing SPDC with $D \sim 0$ is more advantageous compared to ones with $D = 1$ that have been demonstrated in several experiments \cite{Evans:2010jn,YABUNO:2012kk,Jin:2013tx}. 
For the $D = 1$ case, peripheral lobes from the sinc phase-matching function appear along the anti-diagonal ($\Omega_i/\Omega_s = -1$) direction, limiting the single-photon purity to $\sim$85\%. 
For $D \sim  0$ with a broadband pump spectrum, the peripheral lobes are widely distributed along $\Omega_i/\Omega_s \sim  0$, and thus degrade the purity to a lesser extent.

This GVM SPDC source can also achieve high heralding efficiency. 
It has been analyzed \cite{Bennink:2010tc,Ljunggren:2005uj} and demonstrated experimentally \cite{Pereira:2013va,Giustina:2013js,Christensen:2013ux,Dixon:2014bg} that a high heralding efficiency can be realized by using large pump beam waist sizes, relative to the collection modes. 
As discussed in \cite{Guerreiro:2013bk}, roughly speaking, a larger beam waist size has a narrower wavevector distribution, which makes a two-photon state strongly entangled in spatial mode. 
In addition, the collection beam profile of signal and idler modes may be chosen to eliminate spatial-spectral entanglement, i.e., spatial chirp, of each SPDC mode. 
In this case, the signal and conjugate idler spatial mode can each be efficiently coupled into an SMF. 
Our numerical calculations based on the theory described in Appendix B predict this proposed SPDC source can achieve heralding efficiency up to 95\% by optimizing the beam waist sizes of the three interaction modes (see Fig. \ref{predictions} (e)). 

\begin{tiny}
\begin{figure}[t!]
\begin{center}
  \includegraphics[width=0.9\columnwidth,clip]{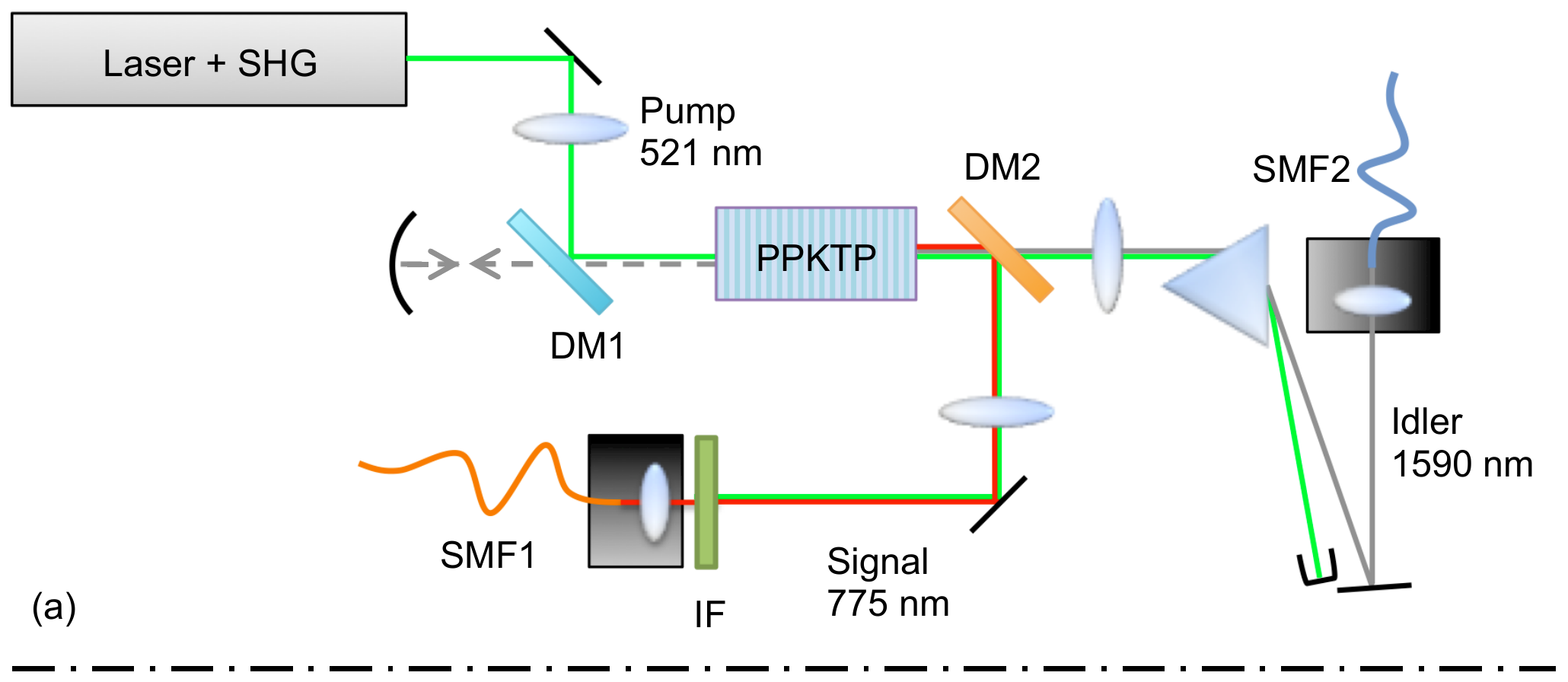}
  \includegraphics[width=0.92\columnwidth,clip]{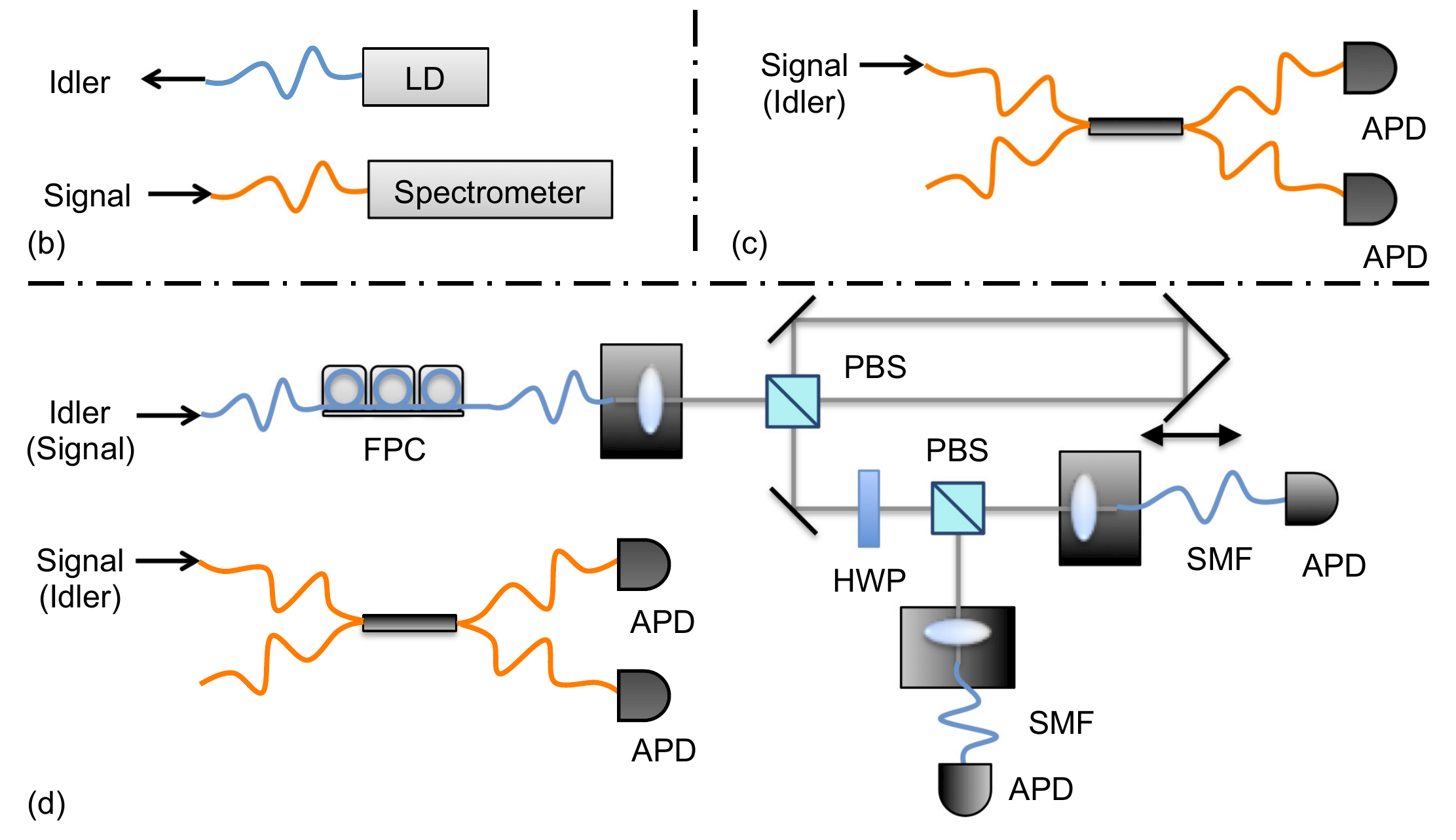}
   \caption{Schematic diagram of experimental setup. (a) Main setup for generation and collection of SPDC photons. (b) Setup for frequency-resolved OPA to reconstruct JSI. Dashed line in (a) is the optical path of the seed laser beam in frequency-resolved OPA measurement. (c) The second-order auto-correlation measurement. (d) Setup for polarization-mode Hong-Ou-Mandel interference experiment.  SHG: second harmonic generation, DM: dichroic mirror, IF: interference filter ($\Delta \lambda =  20 $ nm), SMF: single-mode optical fiber, PBS: polarizing beam-splitter, HWP: half-wave plate, FPC: fiber polarization controller, APD: avalanche photodiode, LD: laser diode. }
\label{setup}
\end{center}
\end{figure}
\end{tiny}

\section{Experiment}

Figure \ref{setup} illustrates a schematic diagram of our experimental setup for generation and characterization of the proposed SPDC and heralded single photons. 
We used the second harmonic (central wavelength 521 nm, pulse duration 160 fs) of a mode-locked Yb-doped fiber laser (pulse repetition rate 100 MHz) as a pump source. 
For the second harmonic generation, we used a lithium tri-borate (LBO) crystal with noncritical phase-matching condition to achieve the wavelength conversion without spatial walk-off. 
The pump beam is focused in the middle of a 20-mm long PPKTP crystal with a poling period of 28.275 $\mu$m, producing signal-idler photon pairs at 775 nm and 1590 nm, respectively. 
Wide-band ($\Delta \lambda$ = 20 nm) interference filters (IF) and a dispersive prism are used for suppression of the pump beam. 
For photon counting measurements we used Si- (InGaAs-) APDs for signal (idler) modes, with detector efficiencies of 60\% (19\%).  
In this SPDC configuration, there is a slight group velocity mismatching between pump and signal modes ($D = -0.024$); we chose this for the given pump wavelength of 521 nm so that the idler wavelength is within the sensitive range of the InGaAs-APDs. 
For the pump wavelength, perfect GVM is satisfied when the signal-idler photon pair is generated at 762 nm and 1647 nm, as shown in Fig. \ref{predictions} (d). 

\section{Fiber coupling and generation efficiency}
According to the numerical simulations shown in Fig. \ref{predictions} (e), we determined a pump beam waist size of 220 $\mu$m and SPDC collection beam waist sizes of 125 $\mu$m and 120 $\mu$m for the signal and idler modes.  
Using the method proposed in \cite{Klyshko:1980uo}, we observed SMF coupling efficiencies of $90\pm 3$\% and $91 \pm 4$\% for the signal and idler modes, respectively$^1$. 
These coupling efficiencies are higher than in previous demonstrations \cite{Pelton:2004uk,Huebel:2007tg,Hentschel:2009ej,Pomarico:2012uj,Stuart:2013kj,Krapick:2013up} using highly nondegenerate SPDC sources, and correspond to the attainable heralding efficiencies assuming lossless collection optics. 

The source brightness is another key metric for an HSPS. 
In our souce we estimate the photon-pair generation efficiency as $1.1 \times 10^4 /\mathrm{ s \cdot mW}$, accounting for the collection optics loss and SMF coupling and SPD efficiencies. 
This value is $\sim$10$\times$ higher than those used for recent 8-photon experiments \cite{Yao:2012fp,Huang:2011eh}, which generate pure single photons by narrowband spectral filtering. 
Although many nondegenerate SPDC sources utilizing the highest $d_{33}$ nonlinear susceptibility have significantly higher brightness \cite{Pelton:2004uk,Huebel:2007tg,Hentschel:2009ej,Pomarico:2012uj,Stuart:2013kj,Krapick:2013up}, those need narrowband spectral filtering to attain state purity, which also significantly reduces brightness and heralding efficiency.

\section{Purity of heralded single photons}
\subsection{JSI measurement}
\begin{tiny}
\begin{figure}[t!]
  \includegraphics[width=1\columnwidth,clip]{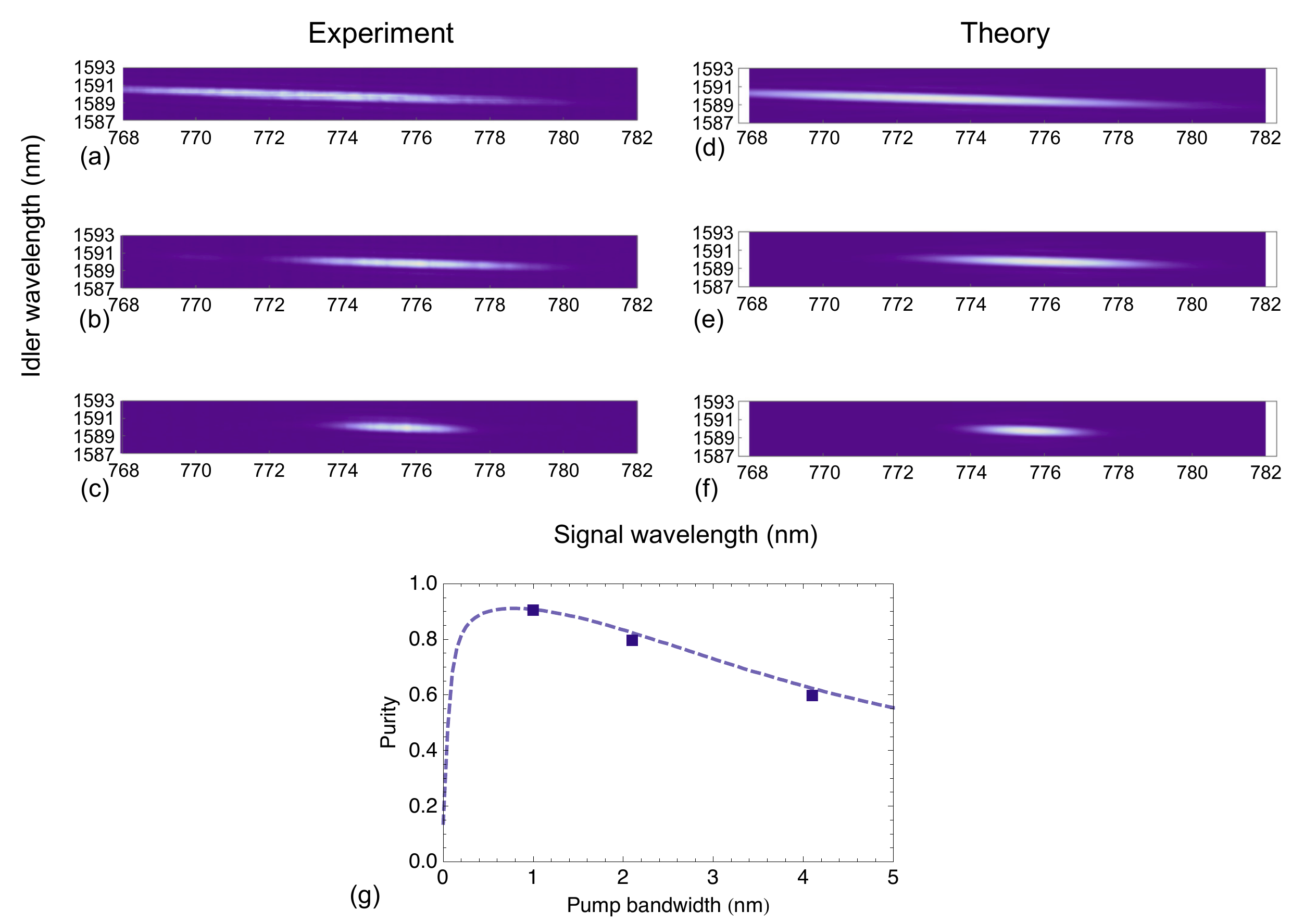}
    \caption{(a-c) Experimentally reconstructed JSI for different pump bandwidth (4.1, 2.1, and 1.0 nm in FWHM). (d-f) Corresponding theoretical prediction for the different pump wavelength and bandwidth. (g) Measured single-photon purity versus pump bandwidth. The dashed curve shows our theoretical predictions.   }
\label{JSI}
\end{figure}
\end{tiny}

In order to characterize the heralded single-photon purity, we performed three different measurements. 
First, in order to map the JSI, i.e., $|f(\omega_s, \omega_i)|^2$ of our SPDC source, we used a frequency-resolved optical parametric amplification (OPA) technique \cite{Liscidini:2013cx,Fang:2014dk,Rozema:2015ey} (see Fig. \ref{setup} (a, b)).
For this measurement, in addition to the pump laser for the SPDC process, we used a laser diode (LD) whose wavelength is stabilized (linewidth $< $ 20 MHz) and tunable around the idler central wavelength. 
The wavelength-stabilized beam is coupled to the collection SMF for the idler mode, propagating backwards along the optical path of the idler photons. 
The backward propagating beam is reflected by a curved mirror and subsequently focused into the PPKTP crystal to seed the OPA process; 
the corresponding signal spectral mode in the JSI is then amplified. 
In other words, only the ``cross-section'' of the JSI at the idler seed wavelength is amplified. 
Thus, the JSI can be reconstructed by measuring the change of the amplified signal spectrum as the seed wavelength is scanned. 
With our 1-mW seed power, the amplification gain over the SPDC process is approximately $10^5\times$, which is sufficient so as to observe the amplified signal spectra by a standard CCD-based spectrometer. 
Compared to the methods relying on coincidence detection of SPDC photon pairs, this OPA-based measurement provides a result much faster and with a higher spectral resolution as determined by the tuning resolution of the seed laser and the spectrometer's resolution. 


Figures \ref{JSI} (a-c) show experimentally reconstructed JSIs for different pump bandwidths; using second-harmonic generation LBO crystals with lengths of 3, 7, and 15 mm, we obtained spectral FWHM bandwidths of 4.1, 2.1, and 1.0 nm, respectively. 
The spectral resolution in the JSI measurement is 0.1 nm for both SPDC modes. 
We find that while signal-mode bandwidth varies according to the pump bandwidth, that of the idler mode remains approximately unchanged. 
This indicates that signal and pump group velocities are well matched, and consequently the phase-matching function has $D \sim 0$. 
We estimate the heralded single-photon purity under the simplifying assumption that all spectral components are in phase, i.e., $f (\omega_s, \omega_i) = \sqrt{|f (\omega_s, \omega_i)|^2}$. 
Computing Schmidt decompositions, we obtain purities of 60.0\%, 79.9\%, and 90.6\% for pump bandwidths of 4.1, 2.1, and 1.0 nm, respectively. 
These estimated purities are in excellent agreement with the theoretical predictions (62\%, 83\%, and 91\% for the same pump bandwidths) shown in Fig. \ref{JSI} (d-f,g). 
Although our observed purity does not reach the attainable value (97\%, predicted in Fig. \ref{predictions} (b-d)) for $D = 0$ because of slightly mismatched group velocities of the pump and signal mode ($D = -0.024$), it is much higher than other sources using nondegenerate SPDC. 
The slightly shifted signal central wavelengths for the different pump bandwidths were due to a temperature change in the LBO crystals (with the tuning  sensitivity of $\sim$0.7 nm/C$^\circ$ around 173 C$^\circ$). 
We also found that the SPDC central wavelengths are insensitive to the temperature change in the PPKTP crystal ($\sim 0.006$ nm/C$^\circ$ for the signal mode from 30 to 100 C$^\circ$); the refractive-index changes of the three SPDC modes are well cancelled to maintain the phase-matching condition. 

\subsection{$g^{(2)} (\tau = 0)$ measurement}
The frequency-resolved OPA can reveal very precise spectral intensity correlations. 
However, it cannot extract ``amplitude'' correlations including spectral phase, because the OPA is a phase-insensitive amplifier for the vacuum field of the signal mode. 
In order to more directly estimate the purity without the spectral-phase assumption, we observed the second-order auto-correlation function at zero time delay $g^{(2)} (\tau = 0)$ for each SPDC mode. 
As described in \cite{Christ:2011ku}, for an SPDC source, $g^{(2)} (\tau = 0) = 1+ P$, where $P$ is the purity of a heralded single photon state. 
For example, $g^{(2)} (\tau = 0) = 2$ for single-mode SPDC and pure heralded single photons, while $g^{(2)} (\tau = 0) = 1$ indicates a Poissonian distribution, i.e., multimode SPDC and heralded single photons in completely mixed states. 
In this experiment, using a fiber-based Hanbury Brown-Twiss setup \cite{Brown:1956vw} (see Fig. \ref{setup} (c)), we measured two APDs' single count rates $S_1$ and $S_2$ and coincidence count rate $C$; 
$g^{(2)} (\tau = 0)$ is then estimated by  
\begin{align}
 g^{(2)} (\tau = 0) = \frac{CR}{S_1 S_2}, 
  \label{g2}
\end{align}
where $R = $ 100 MHz is the pump repetition rate. 
For our SPDC source with an optimized (0.7-nm) pump bandwidth, we observed $g^{(2)} (\tau = 0) = 1.91  \pm 0.02$ $(1.92  \pm 0.01)$ for the idler (signal) mode, corresponding to a heralded single-photon purity of $91\pm2$\% ($92\pm1$\%). 
These purities are consistent with those estimated from the JSI measurement, indicating that the spectral phase of our SPDC source does not degrade the purity.

\subsection{Hong-Ou-Mandel interference measurement}
\begin{tiny}
\begin{figure}[t!]
  \includegraphics[width=1\columnwidth,clip]{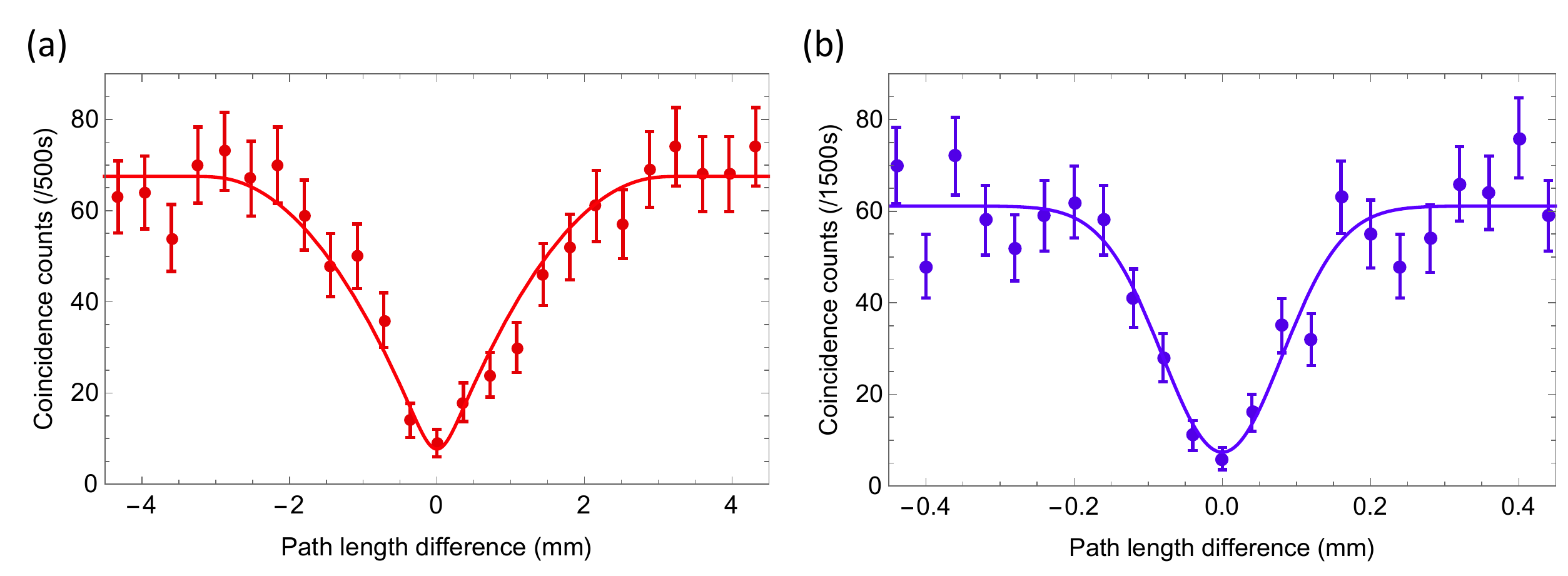}
    \caption{Observed HOMI for (a) sequential idler photons and (b) sequential signal photons.   }
\label{HOMI}
\end{figure}
\end{tiny}

Finally, we measured polarization-mode Hong-Ou-Mandel interference (HOMI) \cite{Hong:1987vi} of heralded single photons generated by sequential pump pulses. 
The HOMI visibility is a direct measure of total state purity (including the effect of the joint spectral phase) and is crucial for many quantum information applications.  
Our setup is shown in Fig. \ref{setup} (d). 
Sequential detections of signal photons herald two idler photons separated by 10 ns, which corresponds to the time interval between two subsequent pump pulses. 
A fiber polarization controller (FPC) adjusts the idler photon's polarization to be a superposition of horizontal ($H$) and vertical ($V$) polarization. 
By using a polarization-dependent optical delay line that delays only $H$-polarized photons for 10 ns, an orthogonally polarized, temporally overlapped two-photon state $\ket{HV}$ is made probabilistically (with 25\% success probability) for the case that the early-born photon with $H$-polarization passes the delay line, and the late-born photon with $V$-polarization is reflected by a polarizing beam-splitter (PBS). 
The two-photon st    ate $\ket{HV}$ is projected onto the diagonal/anti-diagonal ($D/A$) polarization basis ($\ket{D} = (\ket{H} + \ket{V})/\sqrt{2}$, $\ket{A} = (\ket{H} - \ket{V})/\sqrt{2}$), and is then subjected to coincidence detections. 

The results of our HOMI experiment for idler photons are shown in Fig. \ref{HOMI} (a). 
A clearly-observed dip in coincidence counts indicates that the heralded idler photons are in highly pure states. 
The solid curve is the best fit of the experimental data after subtracting background counts (2.4 counts for each data point); we obtained an interference visibility of $91 \pm 4$\% and a dip width of $6.6\pm0.6$ ps. 
As a fitting function, we used a convolution of a Gaussian function and the square of a triangular function, as predicted from the theoretical two-photon joint-spectral amplitude (see Appendix A). 
Using a similar setup, we also observed the HOMI of signal photons heralded by sequential detections of idler photons (see Fig. \ref{HOMI} (b)). 
A longer time of data accumulation was used for the  HOMI measurement involving two signal photons because of less efficient SMF coupling after the delay line. 
By using a Gaussian fit (see Appendix A), the interference visibility and the dip width are estimated to be $90 \pm 5$\% (after subtracting 1.7 background counts) and $0.63\pm0.06$ ps, respectively. 
Although high and similar visibilities are observed for both measurements, as expected from the reconstructed JSI (see Fig. \ref{JSI} (c)), the widths of the two dips differ by a factor greater than 10. 
These observed visibilities are in good agreement with the purities estimated by the JSI and $g^{(2)} (\tau = 0) = 2$ measurements, as well as with our theoretical predictions.

\section{Possible improvements}
It is worthwhile discussing further possible improvements to our HSPS. 
Our observed purity is essentially limited by the slightly mismatched group velocities of pump and signal modes. 
As shown in Fig. \ref{predictions}, for example, perfect GVM is achieved by the pump, signal, and idler wavelengths of $532$ nm, 800 nm, and 1588 nm. 
However, in practice, preparation of the necessary somewhat high-power and broadband optical pulse at $532$ nm would require a bulky pump laser source, e.g., a frequency-doubled optical parametric oscillator pumped by a Ti-Sapphire laser. 
With our current pump wavelength (521 nm), perfect GVM is satisfied by producing signal and idler photons at 762 and 1647 nm, respectively. 
However, it is challenging to characterize photons with wavelengths greater than 1600 nm, for which standard APDs are not sensitive; however, superconducting single-photon detectors tuned for this wavelength may be a viable option. 
Through numerical simulations, we found that eliminating the peripheral lobes of the phase-matching function for our current SPDC source can enhance the purity up to 94\%; modulating poling periods in a KTP crystal can lead to a phasematching function without peripheral sidelobes, e.g. a Gaussian function, instead of the typical sinc-function dependence which does present sidelobes  \cite{Branczyk:2011ti,BenDixon:2013tk}. 
Furthermore, in our SPDC source the poling modulation makes it possible to independently control signal and idler spectra: the signal-mode spectrum is shaped by the pump pulse spectrum, while the idler one can be controlled by the crystal length and poling modulation. 
Finally, we note that because our SPDC source employs collinear phase-matching, it would be compatible with using waveguide structures so as to enhance the photon-pair generation efficiency.

\section{Conclusion}
We have demonstrated the generation of intrinsically pure heralded single photons using a highly nondegenerate, spectrally-factorable SPDC source based on a PPKTP crystal. 
The signal photon at 775 nm permits the use of a Si-APD to herald the idler photon, while the heralded idler photon at 1590 nm is in the telecom L-band, a spectral region where optical fibers are highly transmissive.  A broadband pump source together with matched group velocities of the pump and signal modes produces spectrally factorable photon pairs and pure heralded single photons without the need for spectral filtering. 
With an optimized pump bandwidth, a $\sim$90\% single-photon purity was observed in our JSI,  $g^{(2)} (\tau = 0)$, and HOMI measurements.
We also observed that our source can achieve a high heralding efficiency using appropriate collection optics. 
We anticipate that incorporating the SPDC source with temporal multiplexing techniques \cite{Kaneda:2015dn,migdall:2002hk} will enable near-deterministic single-photon generation to scale up systems for optical quantum information processing. 

\section*{Acknowledgement}
Funding for this work has been provided by NSF Grant No. PHY 12-12439 and PHY 15-20991 and US Army ARO DURIP Grant No. W911NF-12-1-0562.

\section*{Notes}
$^1$In earlier version of the source, we observed slightly asymmetric collection efficiencies for the signal and idler modes. 
We believe this occurred due to non-optimal crystal coatings: we used a dielectric coating on the output facet of the PPKTP crystal so that signal and idler photons escape from the crystal with 99\% efficiency, while 98\% of pump power is reflected on the facet to reduce the power transmitted into the SPDC collection beam paths. 
The input facet was anti-reflection coated (with 99\% transmission) for the pump and idler wavelengths, but not for the signal wavelength.  
With these coatings, however, the reflected pump beam at the output facet can also generate unwanted SPDC in the backward direction. 
The unwanted signal photon is then partially reflected by the non-ideally coated (with $\sim$90\% transmission) input facet of the PPKTP crystal, and as a result, is coupled to a collection SMF with $\sim$2\% coupling efficiency of the main SPDC signal photon. 
This also caused an asymmetry in the second-order auto-correlation function measurement ($g^{(2)} (\tau = 0) = 1.85 \pm 0.01$ and$1.89 \pm 0.02$ for the signal and idler modes), because the unwanted signal photon has a delayed temporal mode (corresponding to the round-trip time inside the PPKTP crystal) relative to the one from the main SPDC, so that the signal mode then has two temporal modes.

\section*{Appendix A: functional form of the HOMI dip}
\setcounter{equation}{0}
\renewcommand{\theequation}{A{\arabic{equation}}}

Here we describe theoretical curves of our observed HOMI dips. 
A schematic of the experiment is shown in Fig. \ref{setup} (a,d). 
We assume that sequentially generated photon pairs have identical JSA $f(\Omega_{s}, \Omega_{i})$, where $\Omega_s$ and $\Omega_i$ are respectively frequency detunings of the signal and idler photons. 
For a given sequentially generated four-photon state $f(\Omega_{s1}, \Omega_{i1})f(\Omega_{s2}, \Omega_{i2})$, the probability of sequentially generated heralded idler photons' coincidence detection, is given by \cite{Ou:2006fz}
\begin{align}
P = \frac{1}{2}(1-\epsilon (\tau) ),
\end{align}
\begin{align}
\epsilon (\tau) =\int d\Omega_{s1} d\Omega_{i1} d\Omega_{s2} d\Omega_{i2} f^*(\Omega_{s1}, \Omega_{i2})f^*(\Omega_{s2}, \Omega_{i1})f(\Omega_{s1}, \Omega_{i1})f(\Omega_{s2}, \Omega_{i2})  e^{-i(\Omega_{i2} - \Omega_{i1})\tau}, 
\label{P_general}
\end{align}
where $\tau$ is the time delay through the optical delay line. 
$\epsilon (\tau)$ denotes the dip function of HOMI. 
Since our SPDC source has the dispersion parameter $D \simeq 0$, the JSA (without normalization factor)
is well approximated by
\begin{align}
f(\Omega_s, \Omega_i) \simeq \exp \left[ -\frac{(\Omega_s + \Omega_i)^2}{\sigma_p^2} \right] \mathrm{sinc} \left[\frac{(k'_p-k'_i)\Omega_i L}{2} \right]. 
\label{S}
\end{align}
Substituting Eq. \eqref{S} into Eq. \eqref{P_general}, we obtain
\begin{align}
\epsilon (\tau) \propto \int dt \exp \left[ -\frac{\sigma_p^2 (\tau- t)^2}{4} \right] T(t, (k'_p-k'_i)L)^2,
\label{dip_idler}
\end{align}
where $T(t, a)$ is a triangular function: 
\begin{align}
T (t, a) =\left\{ \begin{array}{ll}
 a-|t| & (|t|  < a)  \\
0 & (\textrm{otherwise}). \\
\end{array} \right.
\end{align} 
We used Eq.  \eqref{dip_idler}  to find the best fit of our observed HOMI dip shown in Fig. \ref{HOMI} (a). 
For our SPDC source, Eq.  \eqref{dip_idler} more closely follows $T(t, (k'_p-k'_i)L)^2$ which is reflected by the sinc phase-matching function, because the temporal width of the convoluted Gaussian function $2/\sigma_p$ is much narrower than that of the triangular function $(k'_p-k'_i)L$. 
On the other hand, using similar computation, we obtain a dip for the signal photons' interference for $2/\sigma_p \ll (k'_p-k'_i)L$: 
\begin{align}
\epsilon (\tau) \propto \exp \left[ -\frac{\sigma_p^2 \tau^2}{2} \right].  
\end{align}
This Gaussian dip is obtained due to the pump Gaussian spectral envelope function and the group velocity matching of the pump and signal modes. 
Thus, we used a Gaussian function for fitting our observed HOMI of signal photons shown in Fig. \ref{HOMI} (b). 

\section*{Appendix B: fiber coupled SPDC two-photon states}
\setcounter{equation}{0}
\renewcommand{\theequation}{B{\arabic{equation}}}


Following a standard perturbative approach, the quantum state produced by spontaneous SPDC may be expressed as
\begin{equation}
|\Psi\rangle = |{\rm vac}\rangle + \eta \int{\rm d} \vec{k}_s \int{\rm d} \vec{k}_i \,
F(\vec{k}_s,\vec{k}_i)\, \hat{a}_s^\dag(\vec{k}_s) \hat{a}_i^\dag (\vec{k}_i) |{\rm vac}\rangle
\label{twophotStateIn}.
\end{equation}
Here, $F(\vec{k}_s,\vec{k}_i)$ represents the  joint amplitude, and $\eta$ is a constant related to the conversion efficiency. The joint amplitude can in turn be expressed as
\begin{equation}
F(\vec{k}_s,\vec{k}_i)=\alpha(\omega_s+\omega_i)
\phi(\vec{k_s},\vec{k_i}),
\label{E:jointamp}
\end{equation}

\noindent where we have omitted a multiplicative function which is a slow function of the signal and idler frequencies.    The joint amplitude is expressed in terms of the phasematching function 

\begin{eqnarray} \label{phasematchfuct}
\phi(\vec{k}_s,\vec{k}_i) = \exp \left(i \frac{|\underline{k}_\bot|^2}{2k_p} z_0\right)\, \exp \left[i \ts{\frac12} L(k_p+k_{sz}+k_{iz}-\ts{\frac{2\pi}\Lambda})
\right]  \ \exp \left(-\ts{\frac14}w_0^2 |\underline{k}_\bot|^2 \right) {\rm
sinc} \left(\onehalf L \Delta k \right),
\end{eqnarray}

\noindent and of the pump envelope function which, under the assumption of a Gaussian spectral shape, may be written as in Eq. (\ref{pumpenvelope}). 
In the above equations, $w_0$ is the pump beam waist, $L$ is the crystal length, $z_0$ is the position of the pump beam waist relative to the crystal central plane, and  $\Lambda$ is the poling period. 
Note that the phase  in Eq. (\ref{phasematchfuct}) is dependent on the choice of coordinate system; we have placed the origin on the crystal's second face.     Here, we have made the following definitions (an underlined symbol indicates a two-dimensional transverse vector) in terms of the Cartesian components of the signal and idler wavevectors

\begin{eqnarray}
\underline{k}_\bot &\equiv& (k_{sx}+k_{ix},k_{sy}+k_{iy}),\\
\Delta k &\equiv& k_p-\frac{|\underline{k}_\bot|^2}{2k_p}-k_{sz}-k_{iz}+\ts{\frac{2\pi}\Lambda}. \label{Deltak}
\end{eqnarray}
  In writing Eq. (\ref{phasematchfuct}) we have assumed that Poynting vector walkoff for the three interacting fields may be neglected, an assumption which is warranted for a crystal cut parallel to the crystallographic axes, as is the case in our experiment (see above).

The fiber-coupled JSA $f(\omega_s,\omega_i)$ is then given, in terms of $F_{f}(\vec{k_s},\vec{k_i})$ (which represents the joint amplitude taking into account the effects of refraction at the crystal-air interface), as \cite{Vicent:2010ep}

\begin{equation}
\label{PMf}f (\omega_s,\omega_i)=\int\!\!dk_{sx}\int\!\!dk_{sy}\!\!\int \!\!dk_{ix}\int \!\!dk_{iy}\,F_{f}(\vec{k_s},\vec{k_i})\tilde{u}_s^*(\vec{k_s})\tilde{u}_i^*(\vec{k_i}),
\end{equation}

\noindent where  $\tilde{u}_s(\vec{k_s})$ and $\tilde{u}_i(\vec{k_i})$ represent the the fiber-collection modes, expressed as 

\begin{equation}
\label{colectmode}
\tilde{u}_\mu(\vec{k})=\tilde{u}_{0\mu} \, \exp \left(-\ts{\frac14}w_{f\mu}^2[\sec^2\theta_{0\mu}(k_
x-k_\mu \sin \theta_{0\mu})^2+k_{y}^2]\right)\,e^{-i h_\mu}k_{x}.
\end{equation}
In Eq. (\ref{colectmode}) $\mu=s,i$, for the signal(s) and idler (i), $\tilde{u}_{0\mu}$ is a normalization factor, $w_{f\mu}$ represents the beam radius at the beam waist assumed to coincide with the crystal-air interface, $\theta_{0 \mu}$ represents the angular orientation of the collection modes, and $h_\mu$ represents the height of the collection mode on the crystal-air interface. Note that for collinear SPDC, as used in our experiments (see above), $\theta_{0\mu}=0$ and $h_\mu=0$.

From the above expressions, it follows that the two-photon coincidence rate can be calculated as

\begin{equation}
\label{Rc} R_c = \gamma_c\int\!\!d\omega_s\!\!\int \!\!d\omega_i \,|f(\omega_s,\omega_i)|^2.
\end{equation}
Here $\gamma_c$ incorporates various source parameters.  Similarly, the single-photon detection rate (written below for the signal photon; a similar expression exists for the idler photon) can be expressed as

\begin{equation}
\label{Rs} R_s = \gamma_s\int\!\!d\omega_s\!\!\int \!\!d\omega_i \,I_s(\omega_s,\omega_i),
\end{equation}

\noindent where $I_s(\omega_s,\omega_i)$ is the conditioned joint spectral intensity given by

\begin{equation}
\label{CJSI} I_s(\omega_s,\omega_i) = \int\!\!dk_{\bot i}\left|\int \!\!dk_{\bot s}\,F_f(\vec{k_s},\vec{k_i})\tilde{u}_s^*(\vec{k_s})\right|^2.
\end{equation}

For a heralded single-photon source a relevant quantity is the heralding efficiency.  The idler-mode heralding efficiency $\eta_i$ is the probability that a single photon is present in the idler-mode fiber, conditioned on the presence of a single photon in the signal-mode fiber, and is defined by 

\begin{equation}
\label{eta_i} \eta_{i} \equiv \frac{R_c}{R_s},
\end{equation}

\noindent where a corresponding equation may be written for the signal-mode heralding efficiency. In calculating the ratio in Eq. (\ref{eta_i}), the  relationship $\gamma_c/\gamma_s=w_{fi}^2\sec\theta_{0i}$ is useful (with a similar relationship for $\gamma_c/\gamma_i$); the results of this calculation are shown in Fig. \ref{predictions} (e).

\end{document}